\documentclass[conference,final,nofonttune]{IEEEtran}
\IEEEoverridecommandlockouts
\usepackage{cite}
 \usepackage{graphicx}
 \usepackage[table,xcdraw]{xcolor}
\usepackage[normalem]{ulem}
 \useunder{\uline}{\ul}{}
\usepackage{amsmath,amssymb,amsfonts}
\usepackage{microtype}
\usepackage{algorithmic}
\usepackage{graphicx,epstopdf}
\usepackage{textcomp}
\usepackage{subcaption}
\usepackage{boldline} 
\usepackage{epstopdf}
\usepackage{epsfig} 

\usepackage[acronym,nomain]{glossaries}
\makeglossaries
\newacronym{awgn}{AWGN}{additive white Gaussian noise}
\newacronym{vlas}{VLAS}{very large antenna systems}
\newacronym{qam}{16-QAM}{quadrature amplitude modulation}
\newacronym{ofdm}{OFDM}{orthogonal frequency-division multiplexing}
\newacronym{tx}{TX}{transmitter}
\newacronym{bs}{BS}{base station}
\newacronym{ir}{IR}{information receiver}
\newacronym{ap}{AP}{access point}
\glsunset{ap}
\newacronym{cpu}{CPU}{central processing unit}
\newacronym{ezf}{EZF}{eigen-zero-forcing}
\newacronym{csi}{CSI}{channel state information}
\newacronym{svd}{SVD}{singular value decomposition}
\newacronym{iui}{IUI}{inter-user-interference}
\newacronym{bcu}{BCU}{basic component unit}
\glsunset{bcu}
\newacronym{dezf}{DEZF}{decentralized EZF}
\newacronym{gd}{GD}{gradient descent}
\newacronym{admm}{ADMM}{alternating direction method of multipliers}
\newacronym{zf}{ZF}{zero-forcing}
\newacronym{dpc}{DPC}{Dirty Paper Coding}
\newacronym{sca}{SCA}{successive convex approximation}
\newacronym{sinr}{SINR}{signal-to-noise-plus-interference ratio}
\newacronym{sir}{SIR}{signal-to-interference ratio}
\newacronym{snr}{SNR}{signal-to-noise ratio}
\newacronym{wf}{WF}{Wiener filter}
\newacronym{ls}{LS}{least-squares}
\newacronym{ber}{BER}{bit-error rate}
\glsunset{ber}
\newacronym{bers}{BERs}{bit-error rates}
\newacronym{dbp}{DBP}{distributed baseband processing}
\newacronym{cd}{CD}{coordinate descent}
\newacronym{siso}{SISO}{single-input single-output}
\newacronym{mimo}{MIMO}{multiple-input multiple-output}
\newacronym{miso}{MISO}{multiple-input single-output}
\newacronym{simo}{SIMO}{single-input multiple-output}
\newacronym{pd}{PD}{partially decentralized}
\newacronym{fd}{FD}{fully decentralized}
\newacronym{rf}{RF}{radio frequency}
\newacronym{mrt}{MRT}{maximum ratio transmission}
\newacronym{apd}{APD}{approximate partially decentralized}
\newacronym{rv}{RV}{random variable}
\newacronym{iid}{i.i.d.}{independent and identically distributed}
\newacronym{pdf}{pdf}{probability density function}
\newacronym{mu-mimo}{MU-MIMO}{multi-user MIMO}
\newacronym{mmse}{MMSE}{minimum mean-square error}
\newacronym{irc}{IRC}{interference rejection combining}
\newacronym{dl}{DL}{downlink}
\newacronym{tdd}{TDD}{time division duplex}
\newacronym{bcus}{BCUs}{basic component units}
\newacronym{wmmse}{WMMSE}{weighted minimum mean-square error}
\def\BibTeX{{\rm B\kern-.05em{\sc i\kern-.025em b}\kern-.08em
    T\kern-.1667em\lower.7ex\hbox{E}\kern-.125emX}}
\begin{document}
\title{\vspace*{-0.2in}Approximate Partially Decentralized Linear EZF Precoding for Massive MU-MIMO Systems\vspace{-0.0in}}
\author{ \IEEEauthorblockN{Brikena Kaziu\textsuperscript{1}, Nikita Shanin\textsuperscript{1}, Danilo Spano\textsuperscript{2}, Li Wang\textsuperscript{2}, Wolfgang Gerstacker\textsuperscript{1}, and Robert Schober\textsuperscript{1}}
\IEEEauthorblockA{\textit{\textsuperscript{1}Friedrich-Alexander-Universit\"at (FAU) Erlangen-N\"urnberg,  \vspace{-0.18in}}\\
}\\
\IEEEauthorblockA{\textit{\textsuperscript{2}Huawei Technologies Sweden AB\vspace{-0.1in}}\\
}}
\maketitle
\begin{abstract}
Massive multi-user multiple-input multiple-output (MU-MIMO) systems enable high spatial resolution, high spectral efficiency, and improved link reliability compared to traditional MIMO systems due to the large number of antenna elements deployed at the base station (BS). Nevertheless, conventional massive MU-MIMO BS transceiver designs rely on centralized linear precoding algorithms, which entail high interconnect data rates and a prohibitive complexity at the centralized baseband processing unit. In this paper, we consider an MU-MIMO system, where each user device is served with multiple independent data streams in the downlink. To address the aforementioned challenges, we propose a novel decentralized BS architecture, and develop a novel decentralized precoding algorithm based on eigen-zero-forcing (EZF). Our proposed approach relies on parallelizing the baseband processing tasks across multiple antenna clusters at the BS, while minimizing the interconnection requirements between the clusters, and is shown to closely approach the performance of centralized EZF. 

\label{sec:abstract}
\end{abstract}

\vspace*{-0.05in}\section{Introduction}
\begin{figure}[t!]
    \centering
    \includegraphics[scale=0.4,bb = 100 5 274 320]{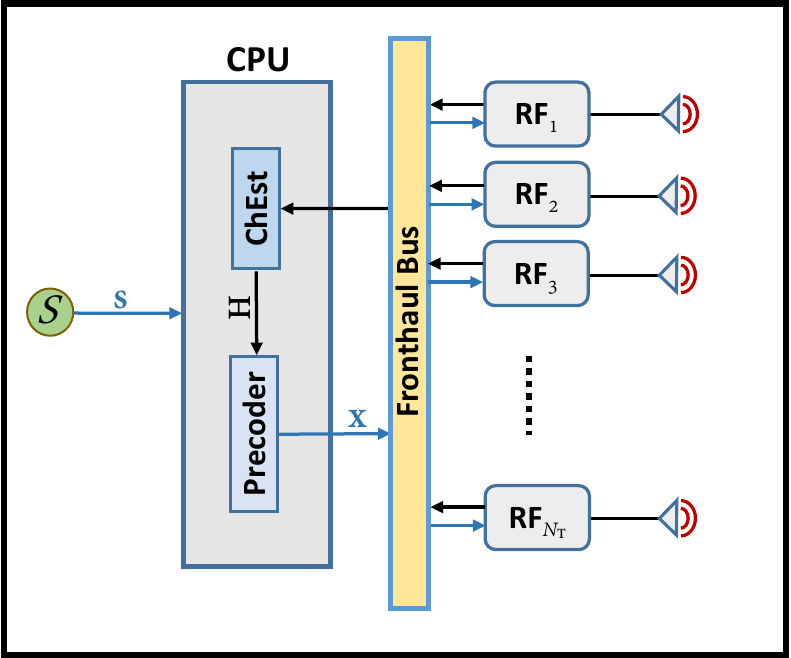}
    \caption{Centralized \gls{bs} architecture with $N_{\mathrm{T}}$ antennas connected via a shared fronthaul bus to a \gls{cpu}.\vspace*{-0.3in}}
    \label{fig:CentalizedArch}
\end{figure}

Massive \gls*{mimo} systems are an extension of traditional \gls*{mimo} communication systems, where hundreds or even thousands of active antenna elements are deployed at the \gls*{bs} and  a large number of user terminals is served simultaneously \cite{Rusek2013,Lu2014,Papa2016,Bjornson2016}. The main advantages of massive \gls{mimo} systems compared to traditional MIMO are a higher spatial resolution, higher spectral efficiency, improved link reliability, and lower latency \cite{Lu2014, Papa2016}. Despite the prominence of massive MIMO, the large number of antenna elements at the BS results also in several implementation challenges \cite{Rusek2013, Yang2013, Lu2014}. In particular, conventional baseband signal processing algorithms that fully exploit the potential of massive MIMO systems, such as \gls{zf} \cite{Hoydis2013} and \gls{ezf} \cite{Sun2010} precoding for single-antenna and multi-antenna users, respectively, rely on a centralized \gls{bs} architecture, see Fig.~\ref{fig:CentalizedArch}. As shown in Fig.~\ref{fig:CentalizedArch}, the \gls{cpu} estimates the \gls{csi} and computes a precoder in the \textit{ChEst}  and \textit{Precoder} blocks, respectively. Additionally, the \gls{cpu} applies the precoding weights to the transmit symbols, which are generated by the source $S$, and forwards the precoded symbols to the \gls{rf} chains. Therefore, for the centralized \gls{bs} design in Fig.~\ref{fig:CentalizedArch}, the complete \gls{csi} and all transmit data streams have to be available at the \gls{cpu} to perform all the baseband processing tasks in a centralized manner. This results in an excessively high amount of raw baseband data, i.e., fronthaul load, that must be exchanged between the \gls{cpu} and the \gls{rf} units via the \textit{fronthaul bus} and increases the computational complexity at the \gls{cpu}.

New distributed baseband processing architectures are proposed in \cite{Li2017, Li2018, Li2019, MLiu2019}, where the antennas at the \gls{bs} are grouped into clusters and equipped with their own circuitry and hardware, facilitating decentralized baseband signal processing. Furthermore, each \gls{bs} cluster locally estimates the corresponding local \gls{csi}. In particular, to compute the local precoding vectors at the antenna clusters, the authors in \cite{Li2017} develop an algorithm, which is based on the alternating direction method of multipliers (ADMM), and relies on iterative consensus-information-sharing between the clusters via a fusion node. The authors in  \cite{Li2018} propose a \gls{pd} algorithm to compute the local precoding vectors at the BS clusters based on the Wiener filter and a feedforward architecture, which achieves exactly the same performance as the centralized counterpart, but with a lower fronthaul overhead. In \cite{Li2018}, the antenna clusters send the Gram matrices computed based on the locally estimated channel matrices to a central node, where the Wiener filter precoding weights are computed. Furthermore, the authors in \cite{Li2018} also propose a \gls{fd} approach, where each antenna cluster determines its local precoding weights using only its local \gls{csi}, without sharing any information with the other clusters. Since the \gls{fd} architecture in \cite{Li2018} does not rely on any information exchange between the BS clusters, it experiences a significant performance loss compared to the \gls{pd} precoding scheme. Based on the \gls{fd} feedforward architecture in \cite{Li2018}, a decentralized \gls{zf} precoding approach using the \gls{cd} algorithm at each \gls{bs} cluster is introduced in \cite{Li2019}. The authors in \cite{MLiu2019} propose a new BS topology referred to as daisy chain architecture, where the antenna units communicate with each other using unidirectional links, and execute the precoding procedure in a sequential fashion. An extension of this work is reported in \cite{Sanchez2020}, where a downlink precoding scheme based on the \gls{cd} algorithm is presented. The results reveal that the required inter-connection data-rate is lower compared to the centralized \gls{zf} approach, however, at the expense of a performance loss.% compared to the \gls{zf}  baseline scheme.
% In contrast to the \gls{PD} precoding scheme, which achieves exactly the same performance as the centralized approach, the \gls{FD} scheme experiences a significant performance loss since the \gls{fd} architecture in \cite{Li2018} does not rely on any information exchange between the BS clusters.

Although the decentralized iterative algorithms in \cite{Li2017, Li2019} achieve almost the same performance as their centralized counterparts, they compute the final precoded symbol vectors instead of the precoder matrices, and therefore, they have to be executed multiple times within a coherence block, making them impractical for systems with a large coherence time, see also \cite{Kaipeng2019}. Furthermore, for their analysis, the authors in \cite{Li2017, Li2018, Li2019, MLiu2019, Sanchez2020} assume that the user devices are equipped with a single antenna and the system load, i.e., the ratio of the total number of users to the total number of antennas at the BS, is lower than 25$\%$, which is an assumption that does not always hold for massive MIMO systems in practice \cite{Bjornson2016}. 

To account for systems with multiple-antenna user devices, the authors in \cite{Zhao2023} show that the mathematical operations of the \gls{ezf} scheme can be distributed among the BS clusters, and thus, the \gls{ezf} precoding vectors can be computed efficiently with a lower fronthaul overhead, while achieving exactly the same performance as the centralized approach. However, the fronthaul load and per-node computational complexities of the \gls{dezf} approach in \cite{Zhao2023} may still be too high for a practical implementation if the number of \gls{bs} clusters and the number of user devices are high. Therefore, we aim at designing a \gls{pd} BS architecture and a linear signal precoding scheme that yield a further fronthaul load and per-node complexity reduction at the \gls{bs}. 

To this end, in this paper, we consider a downlink massive \gls{mu-mimo} system, where the user devices are equipped with multiple antennas and served with multiple data streams. The main contributions of this work are summarized as follows. Motivated by the \gls{pd} architectures in \cite{Li2017,Li2018,Li2019,Zhao2023}, we propose and analyze a novel distributed architecture for massive \gls{mu-mimo} \gls{bs}s. In contrast to the \gls{bs} architectures in \cite{Li2017,Li2018,Li2019,Zhao2023}, the proposed \gls{bs} design does not rely on a \gls{cpu} node, but the computational complexity burden of the baseband signal processing tasks is distributed across different clusters of \gls{bs} antennas. Moreover, for the proposed \gls{bs} architecture, we develop a novel \gls{apd} precoding scheme based on the \gls{ezf} precoding algorithms in \cite{Sun2010,Zhao2023}. We demonstrate that the proposed \gls{apd} approach yields lower fronthaul and per-node computational complexities than the centralized scheme and the \gls{dezf} scheme in \cite{Sun2010} and \cite{Zhao2023}, respectively. Furthermore, our numerical results indicate that the proposed distributed signal processing scheme closely approaches the performance of centralized \gls{ezf}.

The remainder of the paper is organized as follows. In Section II, we introduce the downlink system model and the baseline centralized \gls{ezf} precoding scheme. In Section III, we propose a novel distributed \gls{bs} architecture and the \gls{apd} precoding scheme. In Section IV, we analyze the fronthaul load generated at the \gls{bs} for the centralized \gls{bs} architecture with centralized \gls{ezf} precoding, the proposed decentralized \gls{bs} architecture with \gls{apd} \gls{ezf} precoding, and the \gls{dezf} approach. Furthermore, we numerically evaluate the performance of the proposed scheme and compare the obtained results with the performance of the centralized \gls{ezf}, \gls{dezf}, and \gls{fd} \gls{ezf} precoding schemes. Section V concludes this paper.

\textit{Notations:} Boldface capital letters $\mathbf{A}$ and boldface lower case letters $\mathbf{a}$ denote matrices and vectors, respectively. $A_{i,j}$ stands for the element in the $i$th row and $j$th column of matrix $\mathbf{A}$, whereas $a_i$ is the $i$th element of vector $\mathbf{a}$. $\mathbf{A}^{\mathrm{T}},$ $\mathbf{A}^{\mathrm{H}},$ and $\operatorname{tr}(\mathbf{A})$ denote the transpose, Hermitian transpose, and trace of matrix $\mathbf{A}$, respectively. The Euclidean norm of vector $\mathbf{a}$ and the induced Euclidean norm of matrix $\mathbf{A}$ are represented by $\|\mathbf{a}\|$ and $\|\mathbf{A}\|$, respectively. Furthermore, $\mathbf{A}^{\dagger}$ denotes the Moore-Penrose pseudo-inverse of matrix $\mathbf{A}$. $\mathbb{C}^{m \times n}$ is the set of all $m \times n$ matrices with complex-valued entries and $\mathbf{I}_{N}$ denotes the $N \times N$ identity matrix. $\text{Blkdiag}\{\mathbf{A}_1,\ldots,\mathbf{A}_K\}$ denotes a block diagonal matrix, whose block diagonal entries are $\mathbf{A}_1,\ldots,\mathbf{A}_K$, whereas $\text{diag}\{\mathbf{A}\}$ extracts the main diagonal entries of a matrix $\mathbf{A}$ to a vector. Statistical expectation is represented by $\mathbb{E}\{\cdot\}$.
% and $\mathbf{0}_{N}$ stands for the $N\times 1$ all-zero vector
\label{introduction}
\section{Downlink System Model and Centralized Precoding}
In this section, we introduce the considered \gls{mu-mimo} downlink system model and discuss the centralized \gls{ezf} precoding scheme.
\subsection{Downlink System Model}
We consider a single-cell downlink \gls{mu-mimo} system comprising a \gls{bs} equipped with $N_\mathrm{T}\geq 1$ antennas and $K\geq 1$ user devices. Each user is equipped with $N_\mathrm{R}\geq 1$ antennas and is simultaneously served by the \gls{bs} with $L\leq N_\mathrm{R} $ independent data streams. Thus, the total number of independent data streams broadcast by the BS is given by $L_\mathrm{tot} = KL \leq N_\mathrm{T}$. We assume \gls{tdd} transmission, where channel reciprocity holds. At the beginning of each transmission time interval, the BS computes linear precoding vectors based on the uplink \gls{csi}. Throughout this paper, we assume that perfect \gls{csi}\footnote{In this work, we aim at obtaining an upper bound for the system performance, and therefore, we neglect the impact of imperfect CSI.} of all downlink channels is available at the \gls{bs}, whereas each user device has access only to its own \gls{csi}. The precoding vectors are then applied to the data transmitted to the user devices in the downlink.  

The received complex baseband signal at user $k$, $k=\{1,2,\ldots,K\}$,  can be expressed as follows: \begin{equation}\label{eq:rec_y_k}
    \mathbf{y}_k =  \sqrt{\gamma} \,\mathbf{H}_k\mathbf{W}\mathbf{s} + \mathbf{n}_k,
\end{equation}
 where $\mathbf{W} \in \mathbb{C}^{N_{\mathrm{T}}\times L_{\mathrm{tot}}}$ is the linear precoding matrix at the BS, $\mathbf{H}_k\in \mathbb{C}^{N_\mathrm{R}\times N_\mathrm{T}}$ is the overall downlink channel of user $k$, and $\mathbf{n}_k\in\mathbb{C}^{N_\mathrm{R}\times 1}$ denotes the \gls{awgn} at user $k$, $\forall k$, with zero mean and covariance matrix $\sigma^2_\mathrm{n}\mathbf{I}_{N_\mathrm{R}}$. The independent transmit symbols collected in vector $\mathbf{s}=[s_1\ldots s_{L_{\mathrm{tot}}}]^{\mathrm{T}}\in \mathcal{M}^{L_\mathrm{tot}}$, where $\mathcal{M}$ is the symbol constellation set of the adopted linear modulation scheme, are generated at the source block \textit{S} of the BS, cf. Fig.~\ref{fig:CentalizedArch}. Without loss of generality, in this paper, we assume equal power distribution across all users and transmit data streams. The transmit power loading factor $\gamma$ in \eqref{eq:rec_y_k} is chosen according to $\mathbb{E}\{ \|\sqrt{\gamma}\mathbf{W}\mathbf{s}\|^{2}\} = P_\mathrm{BS}$, where $P_\mathrm{BS}$ is the total power constraint at the BS. 
 
 To detect the current information symbol of its data stream $l$, $l=\{1,2,\ldots,L\}$, user $k$, $\forall{k}$, multiplies the received vector $\mathbf{y}_k$ by the equalization vector $\mathbf{f}_{k, l}\in \mathbb{C}^{N_{\mathrm{R}}\times1}$ as follows:
\begin{align}\label{eq:dec_y_k}
    r_{k,l} &= \mathbf{f}_{k, l}^\mathrm{H} \mathbf{y}_k = \sqrt{\gamma} \, \mathbf{f}_{k, l}^\mathrm{H}\mathbf{H}_k\mathbf{W}\mathbf{s} + \mathbf{f}_{k, l}^\mathrm{H} \mathbf{n}_k \nonumber \\
    & = \sqrt{\gamma} \, \mathbf{c}_{k,l}\mathbf{x} + \tilde{n}_{k,l},
\end{align}
where $\mathbf{x}=\mathbf{W}\mathbf{s}\in \mathbb{C}^{N_\mathrm{T}\times 1}$ is the precoded transmit vector, and $\mathbf{c}_{k,l} = \mathbf{f}_{k,l}^\mathrm{H}\mathbf{H}_k\in \mathbb{C}^{1 \times N_{\mathrm{T}}}$ and $\tilde{n}_{k,l}$ are the equivalent channel vector and the  additive Gaussian noise effective after equalization of stream $l$ at user $k$, $\forall k, l$, respectively. Since we assume perfect \gls{csi} knowledge at the BS and the user devices, user $k$ and the \gls{bs} can compute both the same vector $\mathbf{f}_{k,l}$, $\forall k, l$. Finally, vector $\mathbf{r} = \left[r_{1,1}\;r_{1,2} \ldots r_{k,l} \ldots r_{K,L}\right]^{\mathrm{T}}\in\mathbb{C}^{L_\mathrm{tot}\times 1}$  collecting the equalized data streams of all user devices can be expressed as follows:
\begin{equation}\label{eq:dec_y_k_vec}
    \mathbf{r} = \sqrt{\gamma} \, \mathbf{C}\mathbf{W}\mathbf{s} + \tilde{\mathbf{n}}=\sqrt{\gamma} \, \mathbf{C}\mathbf{x} +  \tilde{\mathbf{n}},
\end{equation}
where $\mathbf{C} = \left[ \mathbf{c}_{1,1}^\mathrm{T} \; \mathbf{c}_{1,2}^\mathrm{T} \ldots \mathbf{c}_{k,l}^\mathrm{T} \ldots \mathbf{c}_{K,L}^\mathrm{T}   \right]^\mathrm{T} \in \mathbb{C}^{L_{\mathrm{tot}}\times N_\mathrm{T}}$ is the equivalent composite downlink channel, and $\tilde{\mathbf{n}} = \left[ \tilde{n}_{1,1} \; \tilde{n}_{1,2} \ldots \tilde{n}_{k,l}\ldots \tilde{n}_{K,L}\right]^\mathrm{T}\in \mathbb{C}^{L_\mathrm{tot}\times 1}$.
\label{sec:sys_model}
\subsection{Centralized \gls{ezf} Precoding}
In the following, we present the linear \gls{ezf} precoding scheme. In addition to performing perfect interference cancellation between different data streams and user devices, linear \gls{ezf} precoding also steers the transmit vector to align with the direction in which the user experiences maximum gain, i.e., where the user has the strongest channel, \cite{Sun2010, Zhao2023}. The optimum direction is determined based on the \gls{svd} of the channel matrix of each user \cite{Sun2010, Zhao2023, matrix85}. To this end, let us first define the \gls{svd} of channel matrix $\mathbf{H}_k$, $\forall k$, as $\mathbf{H}_k = \mathbf{U}_k \mathbf{\Lambda}_k \mathbf{V}_k^{\mathrm{H}}$, where $\mathbf{\Lambda}_k \in \mathbb{C}^{N_{\mathrm{R}}\times N_\mathrm{T}} $ is a diagonal matrix containing the singular values of $\mathbf{H}_k$, i.e., $\lambda_{k, j}$, $j = \{1,2,\ldots, N_\mathrm{R}\}$, in a decreasing order along its main diagonal. Here, $\mathbf{U}_k=\left[\mathbf{u}_{k, 1}~ \mathbf{u}_{k, 2} \ldots \mathbf{u}_{k, N_\mathrm{R}}\right] \in$ $\mathbb{C}^{N_\mathrm{R} \times N_\mathrm{R}}$ and $\mathbf{V}_k=\left[\mathbf{v}_{k, 1}~ \mathbf{v}_{k, 2} \ldots \mathbf{v}_{k, N_\mathrm{T}}\right] \in \mathbb{C}^{N_\mathrm{T} \times N_\mathrm{T}}$ are unitary matrices, where $\mathbf{u}_{k, j}$ and $\mathbf{v}_{k, j}$ are the left and right singular vectors of matrix $\mathbf{H}_k$, respectively, which correspond to the $j$th largest singular value $\lambda_{k, j}$, $\forall k,j$.

To detect the $l$th data stream, $l=\{1,2,\ldots,L\}$, at user $k$, $\forall{k}$, the 
equalization vector $\mathbf{f}_{k,l}$ is chosen by the \gls{cpu} as $\mathbf{f}_{k, l} =\mathbf{u}_{k, l}$ \cite{Sun2010}. Thus, the equivalent channel vector of stream $l$ and user $k$ can be expressed as $\mathbf{c}_{k,l} = \mathbf{u}_{k,l}^\mathrm{H}\mathbf{H}_k$, $\forall k,l$. Then, we determine the precoding matrix $\mathbf{W} = [\mathbf{w}_1 ~ \mathbf{w}_2 \ldots \mathbf{w}_{L_{\mathrm{tot}}}]$, where $\mathbf{w}_i$ is the precoding vector for the $i$th data stream, $i=\{1,\ldots,L_{\mathrm{tot}}\}$, defined as $\mathbf{w}_i = \frac{\hat{\mathbf{c}}_i}{\|\hat{\mathbf{c}}_i\|}$, where $\hat{\mathbf{c}}_i$ is the $i$th column of matrix $\mathbf{C}^\dagger=\mathbf{C}^\mathrm{H} (\mathbf{C} \mathbf{C}^\mathrm{H})^{-1}$ \cite{Sun2010}. Finally, the detected signal vector in \eqref{eq:dec_y_k_vec} can be expressed as:
\begin{equation}\label{eq:dec_y_k_ezf}
    \mathbf{r} = \sqrt{\gamma} \, \tilde{\mathbf{s}} + \tilde{\mathbf{n}},
\end{equation}
where $\tilde{\mathbf{s}}=\left[{s_1}/{\|\hat{\mathbf{c}}_1\|} ~\ldots ~{s_{L_{\mathrm{tot}}}}/{\|\hat{\mathbf{c}}_{L_{\mathrm{tot}}}\|}\right]^{\mathrm{T}}$ is the desired (normalized) symbol vector.

We note that, for the \gls{bs} architecture shown in Fig.~\ref{fig:CentalizedArch}, the computation of the \gls{ezf} precoding vectors requires the full \gls{csi} to be available at the \gls{cpu}. Furthermore, the precoded transmit signal vector $\mathbf{x}$, which scales with $N_\mathrm{T}$, has to be forwarded from the \gls{cpu} to the \gls{rf} chains via the \textit{fronthaul bus}, see Fig.~\ref{fig:CentalizedArch}, resulting in a high fronthaul load for large $N_\mathrm{T}$. Moreover, the \gls{cpu} carries the full computational burden making the centralized implementation of the \gls{ezf} scheme infeasible for large numbers of BS antennas, since complex operations are performed with high-dimensional matrices \cite{Rusek2013,Lu2014}.

\label{sec:CentralizedPrecoding}
\section{Proposed Decentralized Precoding}
To alleviate the high fronthaul requirements and the high computational load of the centralized \gls{ezf} scheme, in the following, we propose a novel distributed \gls{bs} architecture and derive the \gls{apd} \gls{ezf} algorithm.
\subsection{Decentralized \gls{bs} Architecture}
\begin{figure}[t!]
    \centering
    \includegraphics[scale=0.4,bb = 100 2 300 315]{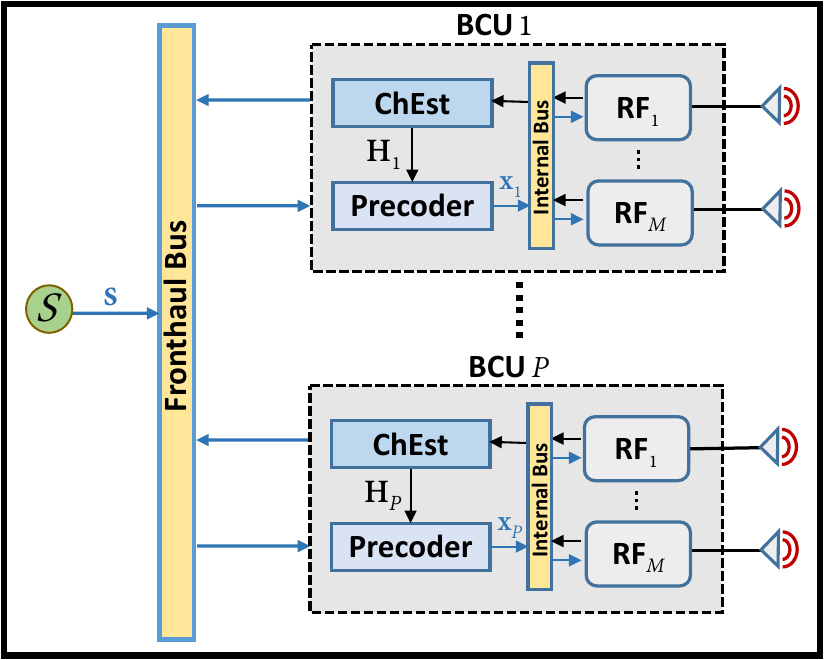}
    \caption{Decentralized \gls{bs} architecture, where the antennas are organized into $P$ \gls{bcu}s, which are connected via a shared fronthaul bus and perform local channel estimation and signal processing.}\vspace*{-0.2in}
    \label{fig:DecentralizedArch}
\end{figure}
 
In contrast to the centralized BS architecture depicted in Fig.~\ref{fig:CentalizedArch}, we propose a decentralized BS architecture as shown in Fig.~\ref{fig:DecentralizedArch}. In particular, we partition the antennas employed at the BS into $P\geq 1$ clusters, which we refer to as basic component units (\gls{bcu}s). Each BCU is equipped with $M$ antennas (i.e., we have $PM= N_\mathrm{T}$), and has its own \textit{ChEst} and \textit{Precoder} block for local \gls{csi} acquisition and local precoder computation, respectively. We note that in contrast to the \gls{bs} architectures in \cite{Li2017,Li2018,Li2019,Zhao2023}, our proposed \gls{bs} design does not employ a \gls{cpu} node for aggregated baseband signal processing. Furthermore, each BCU is connected via the fronthaul bus to the other BCUs, as shown in Fig.~\ref{fig:DecentralizedArch}. Such an architecture enables a scalable \gls{bs} design, where adding or removing antenna elements is equivalent to adding/removing \gls{bcu}s. 

We assume that the $p$th BCU has knowledge of only its own local baseband channel between the corresponding antennas and all deployed user devices. We denote the local channel matrix of BCU $p$ as ${\mathbf{H}}^p = [\mathbf{H}_{1,p}^\mathrm{T}  ~\mathbf{H}_{2,p}^\mathrm{T}  \ldots \mathbf{H}_{K,p}^\mathrm{T} ]^\mathrm{T} \in \mathbb{C}^{KN_\mathrm{R}\times M}$, $p=\{1,2,\ldots,P\}$, where $\mathbf{H}_{k,p}$ is the channel matrix between BCU $p$ and user $k$, $k=\{1,\ldots,K\}$. Based on the estimated local channel matrix $\mathbf{H}^p$ and the information received from the other BCUs, BCU $p$ computes the local precoding matrix $\mathbf{W}_p \in \mathbb{C}^{M\times L_\mathrm{tot}}$, $\forall p$. We note that the complete precoding matrix at the BS can be represented as $\mathbf{W}= [ \mathbf{W}_1^\mathrm{T}~ \mathbf{W}_2^\mathrm{T} \ldots\mathbf{W}_P^\mathrm{T} ]^\mathrm{T}$. Finally, similar to the centralized \gls{bs} design in Fig.~\ref{fig:CentalizedArch}, we assume that the data symbols in vector $\mathbf{s}$ are generated at the source block $S$ of the \gls{bs} and transmitted to the BCUs at the beginning of each transmission time interval, via the \textit{fronthaul bus}, see Fig.~\ref{fig:DecentralizedArch}.

\subsection{Approximate Partially-Decentralized \gls{ezf} Precoding}

In the following, we derive the \gls{apd} \gls{ezf} precoding scheme, which is based on the centralized \gls{ezf} approach discussed in Section \ref{sec:sys_model} and relies on the decentralized architecture depicted in Fig.~\ref{fig:DecentralizedArch}.  

First, let us define block diagonal matrix $\mathbf{D} = \mathrm{Blkdiag}\{\tilde{\mathbf{U}}^\mathrm{H}_1 \ldots \tilde{\mathbf{U}}^\mathrm{H}_K\} \in \mathbb{C} ^{L_{\mathrm{tot}}\times KN_{\mathrm{R}}}$, where $\tilde{\mathbf{U}}_k = [\mathbf{u}_{k,1}\ldots \mathbf{u}_{k,L}]\in \mathbb{C}^{N_\mathrm{R} \times L}$ is the equalization matrix of user $k$, $k=\{1,2,\ldots,K\}$. Next, the equivalent channel matrix $\mathbf{C}$ can be expressed as follows:
\begin{equation}\label{eq:matrix_C}
    \mathbf{C} = \mathbf{D} [{\mathbf{H}}^1 \ldots {\mathbf{H}}^P ]=[\mathbf{C}_1~ \mathbf{C}_2   \ldots\mathbf{C}_P ],
\end{equation}
where $\mathbf{C}_p = \mathbf{D} \mathbf{H}^p$ is the local
equivalent composite channel matrix of \gls{bcu} $p$, $p = \{1,\ldots,P\}$. Then, we compute the Gram matrix $\mathbf{G} = \mathbf{C} \mathbf{C}^\mathrm{H}$ of the equivalent channel matrix as 
\begin{equation}\label{eq:matrix_G}
\mathbf{G} =  \sum_{p=1}^P \mathbf{D} {\mathbf{H}}^p ({\mathbf{H}}^p)^\mathrm{H} \mathbf{D}^\mathrm{H} =  \sum_{p=1}^P \underbrace{\mathbf{C}_p \mathbf{C}_p^\mathrm{H}}_{\mathbf{G}_p}=\sum_{p=1}^P \mathbf{G}_p. 
\end{equation}
Finally, the local precoding matrix can be obtained at \gls{bcu} $p$ as follows:
\begin{equation}\label{eq:matrix_W_p}
    \mathbf{W}_p = [\mathbf{w}_{p,1}\ldots\mathbf{w}_{p,L_\mathrm{tot}}] = \mathbf{C}_p^\mathrm{H} \mathbf{G}^{-1} \mathbf{Q}^{-\frac{1}{2}}, ~\forall p,
\end{equation}
where $\mathbf{w}_{p,i}$ is the local precoding vector at BCU $p$ for data stream $i$, $\forall p, i$, and $\mathbf{Q}$ is a diagonal matrix comprising the squared column norms of matrix $\mathbf{C}^\dagger$ in the main diagonal, i.e., $\mathrm{diag}\{\mathbf{Q}\} = \mathrm{diag}\{(\mathbf{C}^\dagger)^{\mathrm{H}} \mathbf{C}^\dagger\} = \mathrm{diag}\{\mathbf{G}^{-1}\}$. Similar to the \gls{ezf} precoding vectors in Section \ref{sec:CentralizedPrecoding}, the multiplication by $\mathbf{Q}^{-\frac{1}{2}}$ ensures that the columns $\mathbf{w}_{i} = [\mathbf{w}^{\mathrm{T}}_{1,i}\ldots\mathbf{w}^{\mathrm{T}}_{P,i}]^{\mathrm{T}}$, $\forall i$, of the derived composite precoder matrix $\mathbf{W}$ have unit norm.

In order to obtain the local precoding matrices $\mathbf{W}_p$, the BCUs first need to compute matrix $\mathbf{D}$. To this end, as proposed in \cite{Zhao2023}, the BCUs could share the Gram matrices of their local channel matrices $\mathbf{H}^{k,p}_{\mathrm{G}} =\mathbf{H}_{k,p}\mathbf{H}_{k,p}^{\mathrm{H}}$, $\forall k,p$, and then, locally independently compute the \gls{svd} of the aggregate matrix $\mathbf{H}_{\mathrm{G}} = \sum_{p=1}^P \mathbf{H}_{k,p}\mathbf{H}_{k,p}^{\mathrm{H}}$ to obtain the equalization vectors $\mathbf{f}_{k,l}$, $\forall k,l$. However, this approach may yield a high fronthaul load for a large number of user devices $K$. Moreover, in this case, each BCU has to compute the \gls{svd}s of $K$ aggregate Gram matrices, which may lead to undesirably high computational requirements at the \gls{bs}. Therefore, to reduce the fronthaul load and the computational complexity of calculating the precoding vectors, in the following, we approximate matrix $\mathbf{D}$.

First, for each user $k$, $\forall k$, we identify a \textit{strongest} BCU, i.e., the \gls{bcu} with the strongest channel to user $k$, that will lend itself for the computation of the equalization matrix of user $k$. To this end, we introduce the metric $\mathcal{T}_{k,p}=\mathrm{tr}(\mathbf{H}^{k,p}_{\mathrm{G}})$, which is computed at \gls{bcu} $p$, $\forall p,k$. Then, the \gls{bcu}s share the locally computed metrics $\mathcal{T}_{k,p}$, $\forall p,k$, via the \textit{fronthaul bus} and each BCU identifies the \textit{strongest} unit for each user device as follows:
\begin{equation}
{p}^*_k=\underset{{p}=\{1,\ldots,P\}}{\mathrm{argmax} ~{\mathcal{T}}_{k,{p}}}, ~k = \{1, \ldots, K\}.  
\end{equation}
Note that other metrics can be used to define the \textit{strongest} BCU, e.g., $\|{\mathbf{H}}_{k,p}\|$, $\forall k,p$. Nevertheless, since Gram matrix $\mathbf{H}^{k,p}_{\mathrm{G}}$ has to be computed to obtain the block diagonal entries of matrix $\mathbf{G}_p$, cf. \eqref{eq:matrix_G}, we gain in computational efficiency by reusing its values for $\mathcal{T}_{k,p}$, $\forall k,p$, unlike for other possible metrics. Finally, BCU ${p}^*_k$ computes the equalization matrix of user $k$ as $\tilde{\mathbf{U}}_{k,{p}^*_k} = \left[\mathbf{u}^{{p}^*_k}_{k,1} \ldots \mathbf{u}^{{p}^*_k}_{k,L}\right]$, where $\mathbf{u}^{{p}^*_k}_{k,l}$ is the $l$th dominant eigenvector of Gram matrix $\mathbf{H}^{k,p^{_*}}_{\mathrm{G}} = \mathbf{H}_{k,p^{*}}(\mathbf{H}_{k,p^*})^{\mathrm{H}}$, $\forall k$, $l=\{1, \ldots, L\}$.

Next, the \textit{strongest} BCU ${p}^*_k$ for user $k$, broadcasts the locally obtained $\tilde{\mathbf{U}}_{k,{p}^*_k}$, $k = \{1,\ldots,K\}$, to the remaining weaker \gls{bcu}s via the common bus. Thus, matrix $\mathbf{D} = \mathrm{Blkdiag}\{\tilde{\mathbf{U}}_{1,{p}^*_1} \ldots\tilde{\mathbf{U}}_{K,{p}^*_K}\}$ and the local Gram matrix $\mathbf{G}_p$ in \eqref{eq:matrix_G} can be obtained at BCU $p$, $\forall p$. Then, the BCUs share the local Gram matrices $\mathbf{G}_p$, compute the aggregate Gram matrix $\mathbf{G}$ in \eqref{eq:matrix_G}, and calculate the local precoding matrix $\mathbf{W}_p$ in \eqref{eq:matrix_W_p}, for $p=\{1,\ldots,P\}$. Finally, the locally precoded signal vector at BCU $p$, $\forall p$, can be obtained as follows:
\begin{equation}
    \mathbf{x}_p = \mathbf{W}_p \mathbf{s},~ \forall p,
\end{equation}
and is forwarded from the \textit{Precoder} blocks to the \gls{rf} chains via the \textit{internal bus} for downlink transmission, see Fig.~\ref{fig:DecentralizedArch}. 

In contrast to the centralized \gls{ezf} precoding scheme in Section \ref{sec:CentralizedPrecoding} and the \gls{dezf} approach in \cite{Zhao2023}, where the \gls{cpu} node computes an \gls{svd} for $K$ channel matrices, in our scheme, only the \textit{strongest} BCU computes the \gls{svd} of the local channel matrix of a given user. Thus, the computational burden of the \gls{cpu} node of the approach in \cite{Sun2010,Zhao2023} is distributed across different BCUs in the proposed \gls{apd} \gls{ezf} scheme. Furthermore, the dimensionality of the local channel matrices at one BCU depends only on the number of antennas per BCU $M$ and not on the total number of \gls{bs} antennas $N_\mathrm{T}\geq M$. Hence, we conclude that the computational complexity required at one BCU of the proposed scheme is lower compared to the complexity required at the \gls{cpu} node of the scheme in \cite{Sun2010, Zhao2023}.

\label{sec:DecentralizedPrecoding}
\section{Fronthaul Load and Performance Analysis}
In this section, we numerically investigate the fronthaul load and \gls{mu-mimo} system performance of the proposed \gls{apd} \gls{ezf} precoding scheme for the distributed \gls{bs} design in Fig.~\ref{fig:DecentralizedArch}.
\subsection{Fronthaul Load Analysis}
In the following, we numerically evaluate the fronthaul load of the proposed distributed precoding scheme and compare it with the fronthaul load of the centralized \gls{ezf} approach in \cite{Sun2010} and the \gls{dezf} scheme in \cite{Zhao2023}. Let us first denote the fronthaul loads of the centralized \gls{ezf}, the \gls{dezf}, and the proposed \gls{apd} \gls{ezf} schemes by $\zeta_{\mathrm{CEN}}$, $\zeta_{\mathrm{DEZF}}$, and $\zeta_{\mathrm{APD}}$, respectively, which are expressed in terms of the number of real values exchanged via the \textit{fronthaul bus} at the \gls{bs}. Then, the relative fronthaul load gain achieved by the \gls{apd} \gls{ezf} and \gls{dezf} schemes compared to the baseline \gls{ezf} scheme is given by $\hat{\zeta}_{\chi} = 1-(\zeta_{\chi}/\zeta_{\mathrm{CEN}})$, for $\chi = \{\mathrm{APD}, \mathrm{DEZF}\}$.    

In the centralized \gls{ezf} scheme, for the computation of the precoder matrix $\mathbf{W}$, no information is exchanged between the \gls{cpu} and the \gls{rf} chains at the \gls{bs}, given that the precoder is computed in a centralized manner at the \gls{cpu}. For a typical scenario, where the channel is static across $\tau$ consecutive symbol intervals, the only information that is exchanged between the \gls{cpu} and the \gls{rf} chains for each symbol interval is the complex valued precoded signal vector $\mathbf{x}$, see also Fig.~\ref{fig:CentalizedArch}. Thus, the resulting fronthaul load is $\zeta_{\mathrm{CEN}}= 2\tau N_\mathrm{T}$. Meanwhile, in the proposed the APD scheme, the information shared between BCUs comprises symbol vector $\mathbf{s}$, metrics $\mathcal{T}_{k,p}$, and matrices $\Tilde{\mathbf{U}}_{k,p_k^*}$ and $\mathbf{G}_p$, respectively, which results in $\zeta_{\mathrm{APD}} = PK + 2L_{\mathrm{tot}}N_{\mathrm{R}}-L_{\mathrm{tot}} + PL_{\mathrm{tot}}^2+2\tau L_{\mathrm{tot}}$. Lastly, the fronthaul load of the \gls{dezf} scheme, $\zeta_{\mathrm{DEZF}}$, is computed as in \cite{Zhao2023}. 

% $\zeta_{\mathrm{DEZF}} = PKN^2_{\mathrm{R}} + 2L_{\mathrm{tot}}N_{\mathrm{R}}+ P+1)L_{\mathrm{tot}}^2$.
%exchanged between the BCUs for the computation of the local precoder matrices takes place
%In contrast to the centralized approach, where the precoded symbol vector $\mathbf{x}$ has to be forwarded from \gls{cpu} to \gls{rf} chains for each symbol interval, for the APD scheme, the fronthaul load $\zeta_{\mathrm{APD}}$ in the \textit{fronthaul bus} is generated only once for $\tau$ consecutive symbol intervals. 
\begin{table}[t!]
\caption{Fronthaul load gain of the \gls{apd} \gls{ezf} and \gls{dezf} schemes, for varying system load $\eta$.}\vspace*{-0.06in}
\label{tab:table1}
\resizebox{\columnwidth}{!}{%
\begin{tabular}{cccccc}
\hlineB{3}
\multicolumn{1}{V{3}c V{3}}{\textbf{System   Parameters}} & \multicolumn{1}{c|}{$M = 64$} & \multicolumn{1}{c|}{$P = 4$} & \multicolumn{1}{c|}{$L = 2$} & \multicolumn{1}{c|}{$N_\mathrm{R} =   4$} & \multicolumn{1}{c V{3}}{$\tau = 65$} \\ \hline \hline
\multicolumn{6}{V{3}c V{3}}{\textbf{Fronthaul load gain}} \\ \hlineB{3}
\multicolumn{2}{V{3}c V{3}}{$K, ~\eta$} & \multicolumn{2}{c V{3}}{\cellcolor[HTML]{C6EFCE}$\hat{\zeta}_{\mathrm{APD}}$ (proposed)} & \multicolumn{2}{c V{3}}{\cellcolor[HTML]{B4C6E7}$\hat{\zeta}_{\mathrm{DEZF}}$} \\ \hlineB{3}
\multicolumn{2}{V{3}c V{3}}{$K = 16, ~\eta =   12.50\%$} & \multicolumn{2}{c V{3}}{74.33\%} & \multicolumn{2}{c V{3}}{68.27\%} \\ \hline
\multicolumn{2}{V{3}c V{3}}{$K = 24, ~\eta =   18.75\%$} & \multicolumn{2}{c V{3}}{52.26\%} & \multicolumn{2}{c V{3}}{40.87\%} \\ \hline
\multicolumn{2}{V{3}c V{3}}{$K = 32,~ \eta =   25.00\%$} & \multicolumn{2}{c V{3}}{24.04\%} & \multicolumn{2}{c V{3}}{5.77\%} \\ \hline
\multicolumn{2}{V{3}c V{3}}{$~K =36, ~\eta   = 28.125\%$} & \multicolumn{2}{c V{3}}{7.62\%} & \multicolumn{2}{c V{3}}{{\color[HTML]{FF0000} -14.66\%}} \\ \hlineB{3}\vspace*{-0.25in}
\end{tabular}
}
\end{table}
\begin{table}[t!]
\caption{Fronthaul load gain of the \gls{apd} \gls{ezf} and \gls{dezf} schemes, for varying number of BCUs $P$.}\vspace*{-0.06in}
\label{tab:table2}
\resizebox{\columnwidth}{!}{%
\begin{tabular}{cccccc}
\hlineB{3}
\multicolumn{1}{V{3}c V{3}}{\textbf{System  Parameters}} & \multicolumn{1}{c|}{$N_{\mathrm{T}} = 256$} & \multicolumn{1}{c|}{$K = 16$} & \multicolumn{1}{c|}{$L = 2$} & \multicolumn{1}{c|}{$N_\mathrm{R} =   4$} & \multicolumn{1}{c V{3}}{$\tau = 65$} \\ \hline \hline
\multicolumn{6}{V{3}c V{3}}{\textbf{Fronthaul load gain}} \\ \hlineB{3}
\multicolumn{2}{V{3}c V{3}}{$P, ~M$} & \multicolumn{2}{c V{3}}{\cellcolor[HTML]{C6EFCE}$\hat{\zeta}_{\mathrm{APD}}$ (proposed)} & \multicolumn{2}{c V{3}}{\cellcolor[HTML]{B4C6E7}$\hat{\zeta}_{\mathrm{DEZF}}$} \\ \hlineB{3}
\multicolumn{2}{V{3}c V{3}}{$P = 4, ~M = 64$} & \multicolumn{2}{c V{3}}{74.33\%} & \multicolumn{2}{c V{3}}{68.27\%} \\ \hline
\multicolumn{2}{V{3}c V{3}}{$P = 8, ~M = 32$} & \multicolumn{2}{c V{3}}{61.83\%} & \multicolumn{2}{c V{3}}{52.88\%} \\ \hline
\multicolumn{2}{V{3}c V{3}}{$P = 16, ~M = 16$} & \multicolumn{2}{c V{3}}{36.83\%} & \multicolumn{2}{c V{3}}{22.12\%} \\ \hlineB{3} \vspace*{-0.4in}
\end{tabular}
}
\end{table}

Next, we define the system load $\eta$ as the ratio between the total number of transmit data streams, $L_{\mathrm{tot}}$, and the total number of transmit antennas at the \gls{bs}, $N_{\mathrm{T}}$, i.e., $\eta = L_{\mathrm{tot}}/N_{\mathrm{T}}$. To analyze the impact of $\eta$ on the fronthaul load gain of the \gls{apd} \gls{ezf} and \gls{dezf} schemes, we fix the remaining system parameters, see Table \ref{tab:table1}. From the results shown in Table \ref{tab:table1}, we observe that the proposed \gls{apd} \gls{ezf} scheme outperforms both the centralized \gls{ezf} and the \gls{dezf} approach in terms of fronthaul complexity requirements, i.e., $\hat{\zeta}_{\mathrm{APD}}> 0$ and $\hat{\zeta}_{\mathrm{APD}}> \hat{\zeta}_{\mathrm{DEZF}}$, respectively, for all considered cases. In particular, for $\eta = 12.5\%$, the \gls{apd} \gls{ezf} scheme achieves a gain of $74.33\%$ compared to centralized \gls{ezf}, nearly $6\%$ more than the gain achieved by the \gls{dezf} approach. Furthermore, as the system load increases, the gap between the \gls{apd} \gls{ezf} and the \gls{dezf} schemes becomes more significant, exceeding $20\%$ for $K = 36$. Additionally, for $K = 36$, in contrast to our proposed scheme, which yields a positive gain of $7.62\%$, \gls{dezf} exceeds the fronthaul load of centralized \gls{ezf}, i.e., $\hat{\zeta}_{\mathrm{DEZF}}<0$. Finally, we note that an increase of the system load leads to lower fronthaul load gains compared to centralized \gls{ezf} for both the \gls{apd} \gls{ezf} and the \gls{dezf} schemes given that their fronthaul loads depend on the total number of users $K$. 

Next, we consider the impact of the number of antennas per BCU, $M$, on the fronthaul load gain $\hat{\zeta}_{\chi}$, $\forall \chi$. Similar to the analysis conducted with respect to $\eta$, we fix the remaining system parameters and vary $M$. As the results given in Table \ref{tab:table2} show, the proposed \gls{apd} \gls{ezf} scheme entails a lower fronthaul load compared to centralized \gls{ezf} and \gls{dezf}, i.e., $\hat{\zeta}_{\mathrm{APD}}> 0$ and $\hat{\zeta}_{\mathrm{APD}}> \hat{\zeta}_{\mathrm{DEZF}}$, respectively, for all considered scenarios. Furthermore, the benefits of the \gls{apd} \gls{ezf} scheme over the \gls{dezf} approach become more prominent as the number of antennas per BCU $M$ decreases. In particular, for $(P, M)=(16,16)$, our scheme outperforms \gls{dezf} by more than $14\%$. Although a lower $M$, i.e., higher number of BCUs $P$ at the \gls{bs}, enables a lower computational complexity per BCU, see Section \ref{sec:DecentralizedPrecoding}, it results in a higher fronthaul load for the \gls{apd} \gls{ezf} and \gls{dezf} approaches, i.e., we obtain lower $\hat{\zeta}_{\chi}$, $\forall \chi$, see Table \ref{tab:table2}. Thus, a more modular \gls{bs} design, i.e., lower $M$, poses a trade-off between fronthaul load and  computational complexity.

\label{sec:front_comput}
\subsection{Performance Analysis}
In this section, we numerically evaluate the performance of the proposed precoding scheme. To obtain channel matrices $\mathbf{H}_{k}$, $\forall k$, for realistic communication scenarios, we set the carrier frequency $f_c$ to 2.1 GHz and model the communication environment using QuaDRiGa \cite{Quad2021}. The BCUs at the \gls{bs} are modeled as uniform planar arrays (UPAs) with cross-polarized antennas. The spacing between any two adjacent antenna elements at the \gls{bs} is set to 0.5$\lambda$, where $\lambda = \frac{c}{f_{\mathrm{c}}}$ is the wavelength and $c$ is the speed of light. The users are placed randomly in a 120° sector in a range of 50~m-120~m from the BS and are equipped with $N_{\mathrm{R}}\times 1$ vertical uniform linear arrays (ULAs) with omni-directional antennas. The transmit symbols are chosen from a 16-ary \gls{qam} constellation. Throughout this section, the number of antennas at the \gls{bs} is $N_{\mathrm{T}} = 256$, the number of streams per user is $L=2$, and the number of antennas per user equals $N_\mathrm{R}=4$.

In the following, we compare the uncoded \gls{bers} obtained for the proposed \gls{apd} \gls{ezf} scheme with the corresponding results for the centralized \gls{ezf}/\gls{dezf} \cite{Sun2010, Zhao2023} and the FD \gls{ezf} \cite{Li2018} schemes, respectively. For the FD \gls{ezf} scheme, we optimistically assume that the equalization vectors $\mathbf{u}_{k,l}$, $\forall k,l$, are known to all the BCUs, and BCU $p$ computes the local precoding vector of data stream $i$ as $\mathbf{w}_{p,i} = \frac{\hat{\mathbf{c}}_{p,i}}{\|\hat{\mathbf{c}}_{p,i}\|\sqrt{P}}$, where $\hat{\mathbf{c}}_{p,i}$ is the $i$th column of $ \mathbf{C}_p^{\dagger}$, $\forall p, i$, \cite{Li2018}.

\begin{figure}[t!]
    \centering
   \hspace*{+0.3in}\includegraphics[width=1.7\linewidth]{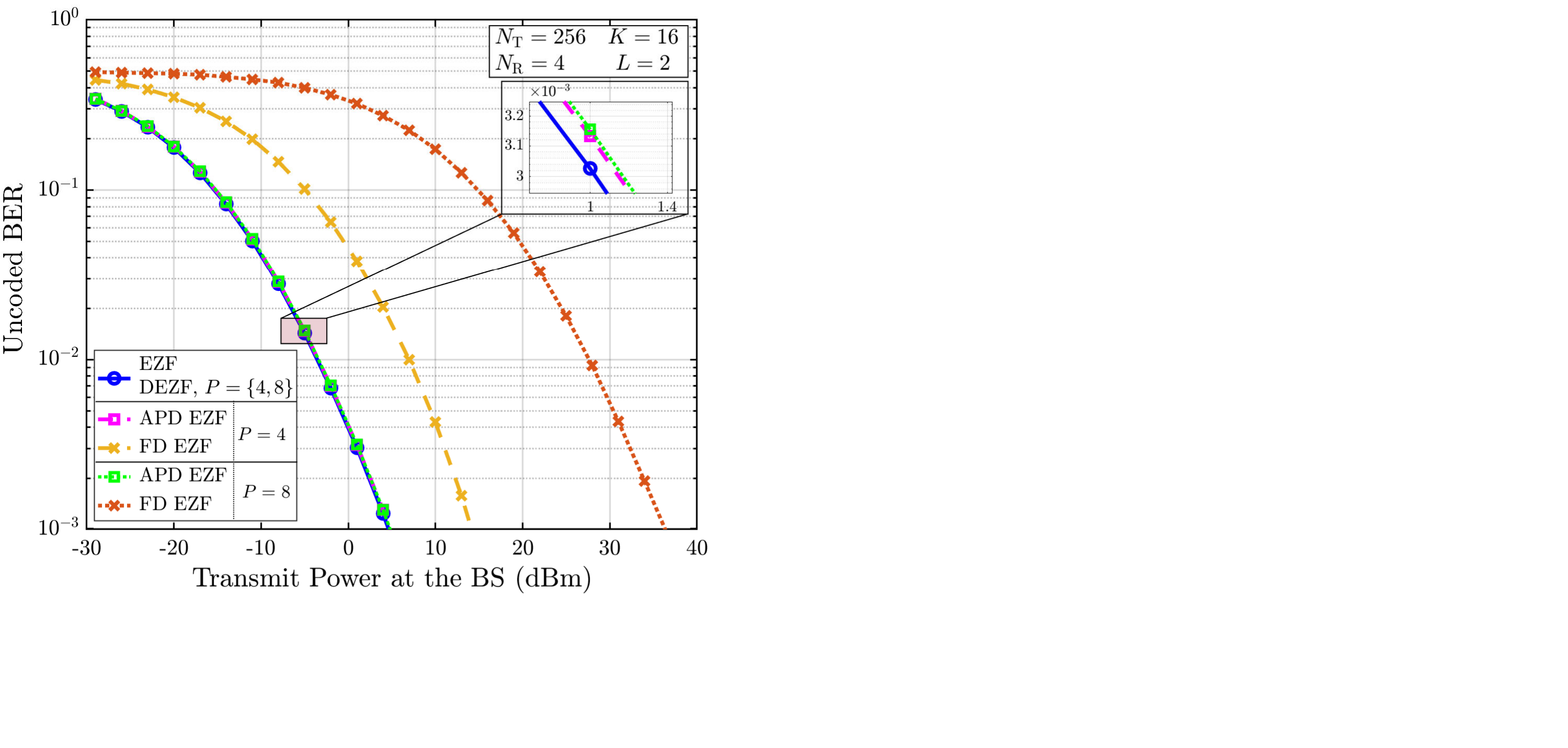}\vspace*{-0.6in}
    \caption{Uncoded BER versus trasmit power at the BS for varying number of BCUs $P$.} \vspace*{-0.2in}
    \label{fig:uncoded_BER_diff_P}
\end{figure}

To analyze the impact of the number of antennas per BCU $M$ on the uncoded \gls{ber} of the proposed \gls{apd} \gls{ezf} scheme, we fix the number of users to $K=16$, and consider $(P,M)=\{(4,64),(8,32)\}$. As the results in Fig.~\ref{fig:uncoded_BER_diff_P} show, the proposed \gls{apd} \gls{ezf} scheme achieves almost the same performance as centralized \gls{ezf} and \gls{dezf}, for all considered cases. Furthermore, the \gls{apd} \gls{ezf} approach significantly outperforms the \gls{fd} \gls{ezf} scheme for all considered $(P,M)$ combinations. For the \gls{fd} \gls{ezf} scheme, we still serve all  $K$ users, but local \gls{ezf} based on $M\leq N_\mathrm{T}$ \gls{bs} antennas instead of $N_\mathrm{T}$ antennas is performed \cite{Li2018}, and therefore, a decrease of $M$ has a detrimental effect on the performance of the \gls{fd} \gls{ezf} scheme. For example, for $(P,M)=(4,64)$, the \gls{fd} \gls{ezf} scheme experiences a loss of approximately 9 dB in combining gain compared to the \gls{apd} \gls{ezf} scheme, and the loss increases further to nearly 30 dB for $(P,M)=(8,32)$. In contrast, a more modular \gls{bs} architecture, i.e., smaller $M$ and larger $P$, does not compromise the performance of the proposed \gls{apd} \gls{ezf} approach. 

\begin{figure}[t!]
    \centering
   \hspace*{+0.2in}\includegraphics[width=1.7\linewidth]{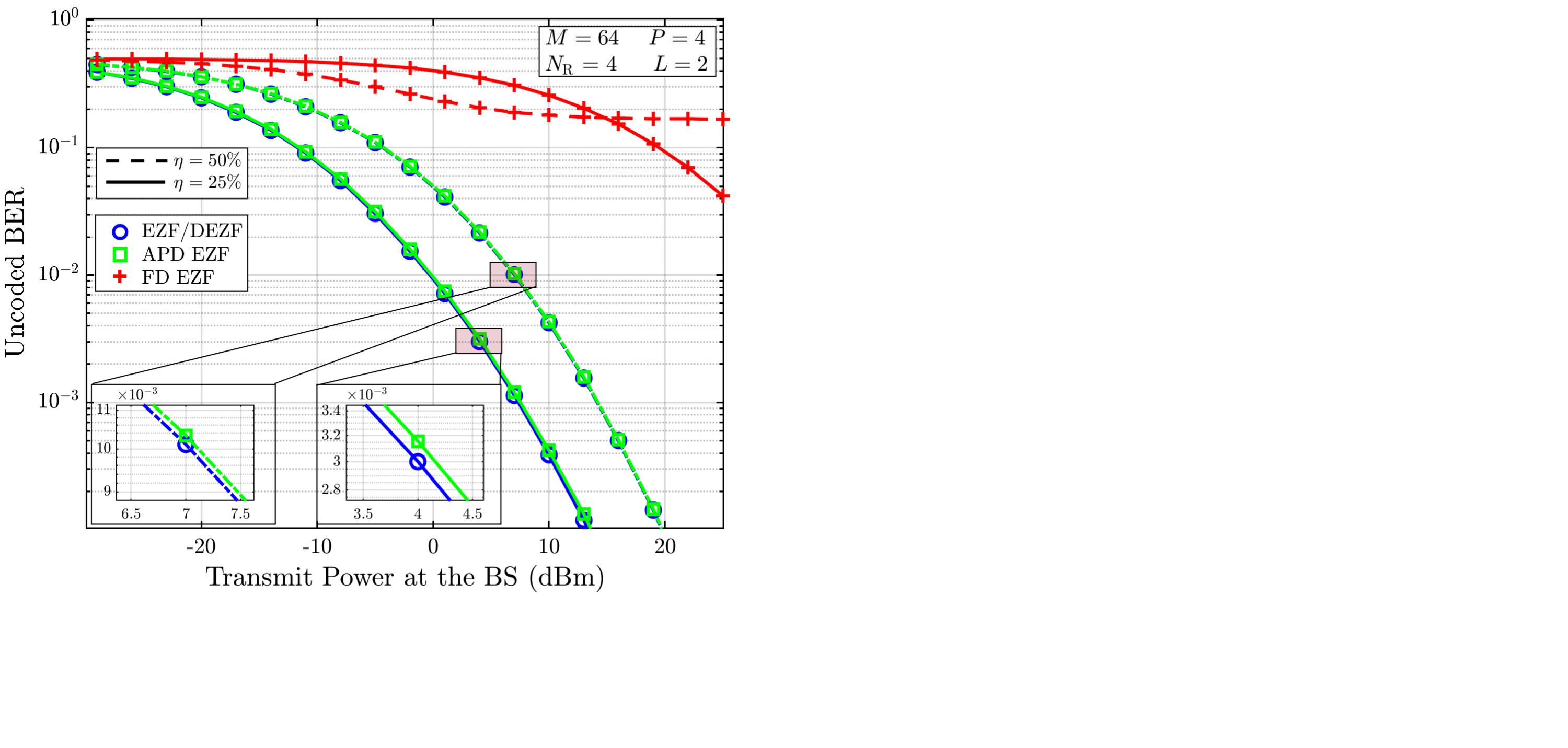}\vspace*{-0.6in}
    \caption{Uncoded BER versus transmit power at the BS for varying system load $\eta$.}\vspace*{-0.208in}
    \label{fig:uncoded_BER_diff_loads}
\end{figure}

Next, we study the impact of the system load $\eta$ on the \gls{ber} performance of the proposed \gls{apd} \gls{ezf} scheme for $P =4$ and $K = \{32,64\}$, i.e., $\eta=\{25\%, 50\% \}$. As the results depicted in Fig.~\ref{fig:uncoded_BER_diff_loads} show, there is almost no gap between the \gls{ber} for the proposed \gls{apd} \gls{ezf} and the centralized \gls{ezf} and \gls{dezf} schemes, for all considered values of $\eta$. Furthermore, the proposed scheme significantly outperforms the \gls{fd} approach. In fact, in contrast to the proposed approach, the \gls{fd} \gls{ezf} scheme is not able to cancel the \gls{iui} for $\eta = 50\%$ since $M < L_\mathrm{tot}$, which leads to an error floor.

For both of the cases considered above, i.e., higher $\eta$ and $P$, respectively, the proposed \gls{apd} \gls{ezf} scheme still performs channel inversion for perfect \gls{iui} cancellation. Nevertheless, \gls{apd} \gls{ezf} does not perform beam steering using the strongest eigenchannels for each user. Instead, it selects sub-optimal channels based only on the eigenbeams of the local \gls{csi} of the \textit{strongest} BCU. As the BER curves in Fig.~\ref{fig:uncoded_BER_diff_P} and Fig.~\ref{fig:uncoded_BER_diff_loads} indicate, the \textit{strongest} BCU is capable of selecting close-to-optimal channels for $M>L$, resulting in only a negligible performance loss compared to the centralized \gls{ezf}/\gls{dezf} schemes.
\label{sec:BER}
\section{Conclusions} In this paper, we have proposed a novel approach for distributed \gls{bs} signal processing in massive downlink multiple-antenna \gls{mu-mimo} systems, where each user device is simultaneously served with multiple data streams. We have proposed a novel decentralized BS architecture which does not rely on a \gls{cpu} node for centralized baseband signal processing. Instead, it distributes the computational complexity burden of the baseband signal processing tasks across different clusters of BS antennas. Next, for the proposed BS architecture, we have developed a novel decentralized precoding algorithm based on linear EZF. Our numerical simulations showed that the proposed \gls{apd} \gls{ezf} scheme can achieve a significantly lower fronthaul load compared to the baseline centralized \gls{ezf} and \gls{dezf} schemes, for various numbers of BS antenna clusters and different values of system load. Furthermore, we have shown that the \gls{apd} \gls{ezf} scheme achieves nearly the same \gls{ber} performance as the considered baseline \gls{ezf} and \gls{dezf} schemes and it outperforms the \gls{fd} \gls{ezf} approach for different system setups.%, demonstrating the efficiency of the proposed algorithm.

\label{sec:conclusions}
\bibliographystyle{IEEEtran}
\bibliography{main}
\end{document}